\documentclass[]{aa}
\usepackage[]{psfig}
\usepackage{txfonts}
\begin{document}

\title{
A photometric monitoring of bright high-amplitude $\delta$ Scuti stars}
\subtitle{I. The double-mode pulsation of V567~Ophiuchi}

\author{L. L. Kiss\inst{1,2} \and A. Derekas\inst{1} \and 
Sz. M\'esz\'aros\inst{1,3} \and P. Sz\'ekely\inst{1}}

\institute{Department of Experimental Physics and Astronomical Observatory,
University of Szeged,
Szeged, D\'om t\'er 9., H-6720 Hungary
\and
Hungarian E\"otv\"os Fellowship, Instito de Astrof\'\i sica de Andaluc\'\i a,
CSIC, Granada, Spain
\and Department of Optics and Quantum Electronics, University of Szeged,
POB 406, Szeged, H-6701 Hungary}

\titlerunning{V567~Ophiuchi, a double-mode pulsator}
\authorrunning{L. L. Kiss et al.}
\offprints{l.kiss@physx.u-szeged.hu}
\date{}

\abstract{
We present the first results of an observational project, which
addresses the period changing behaviour of a sample of  
high-amplitude $\delta$ Scuti stars. In this paper we discuss 
the double-mode nature of V567~Ophiuchi. It was
observed on 15 nights in two consecutive years in order to resolve
the long-standing ambiguity related to its secondary period. 
A frequency analysis of almost 5000 individual single-filtered 
CCD $V$ measurements resulted in 
two independent frequencies ($f_1=6.6879$ d$^{-1}$ and
$f_2=11.8266$ d$^{-1}$) with a ratio of $f_1/f_2=0.565$. 
Earlier data taken from the literature were used to refine the dominant
period, and the re-analysis supports the existence of the
secondary period. Possible asteroseismological implications
are briefly discussed.
\keywords{stars: variables: general -- stars: oscillations --
$\delta$ Sct -- stars: individual: V567~Oph}}
 
\maketitle

\section{Introduction}

High-amplitude $\delta$ Scuti stars (hereafter HADS) form an interesting
subgroup of short-period pulsators located inside the classical
instability strip near the main-sequence. Their light variation
is characterized by relatively large amplitudes
(a conventional limit is $A_{\rm V}\geq0\fm30$) which is associated
with one or two stable frequencies. The variability of these 
stars has been interpreted as caused by radial pulsation in fundamental or
low-order radial overtone pulsation (Rodr\'\i guez et al. 1996, 
Petersen \& Christensen-Dalsgaard 1996), although some 
empirical evidence is present for microvariability due to weakly
excited high-order radial or non-radial modes of pulsation 
(Garrido \& Rodr\'\i guez 1996). It also appears that they exhibit 
a period-luminosity relation which has been 
studied by several authors (e.g. McNamara 1997, Petersen \& H\o g 1998).
Previous studies based on bright field stars have recently been supplemented 
by analyses of tens of objects discovered by the MACHO
and OGLE projects (Alcock et al. 2000, Poretti 2001).

Breger \& Pamyatnykh (1998) tried to infer evolutionary conclusions
from long-term observational records of selected HADS. Although they 
found evidence of period changes in a significant fraction of HADS, it 
was not possible to relate them to stellar evolution. The detected
period changing behaviours range from continuous period decrease to
continuous period increase, and period jumps might also be
present in several stars. Furthermore, cyclic period variations
due to possible binary light-time effect were inferred in a few cases
(see references in Breger \& Pamyatnykh 1998). 
However, despite the large amplitudes, short periods and moderate 
brightnesses of the stars studied, many of them need period updates
as the latest observations
in the literature were obtained almost two decades ago. That is why we 
started a photometric monitoring of bright northern HADS. 
Our target stars
include all HADS brigther than $V\sim11^{\rm m}$ situated in 
favourable positions in the northern sky  
(our sample is partially overlapping 
with that of Breger \& Pamyatnikh). The observations started in 1995 
(Kiss \& Szatm\'ary 1995), and since then
we have obtained unfiltered (one star), Johnson (eight stars) and 
Str\"omgren (one star) photometric observations for ten stars in order 
to get an updated view of their period changes (some early results have
already been published in Kiss \& Szatm\'ary 1995 and Kiss \& Derekas 2000).
Here we report on results for V567~Ophiuchi,
which is the only one that clearly shows
double-mode pulsation.
The period changes of the remaining stars will be discussed in a 
companion paper.

The light variations of V567~Oph (=BD+1$^\circ$3547, 
$\langle V\rangle\approx 11\fm2$,
$A_{\rm V}=0\fm34$, $P=0.1495$ d, spectral type A6--F1, Powell et al. 1990)
were discovered by Hoffmeister (1943) giving a period ($\approx 1/8$ d) 
that was an alias of the true one ($\approx 1/7$ d). De Bruyn (1972) 
was the first who determined the period accurately. 
The observational record until 1990 is quite numerous and has been
summarized in Powell et al. (1990). These authors carried out a detailed
photometric and spectroscopic study of V567~Oph, which was based on
Str\"omgren photometry and medium-resolution optical spectroscopy. Besides
determining the fundamental physical parameters of the star, Powell et al.
(1990) suspected the existence of a secondary period, although they could not
draw a firm conclusion on it. In the same year, Poretti et al. (1990) 
presented Fourier decomposition of three HADS. The true period of 0.149 d 
was established for V567~Oph and they concluded that it was monoperiodic. 
This conclusion was
based on four nights of observations distributed in two separate years (1984 and
1986). Rodr\'\i guez et al. (1996) studied the phase shifts and amplitude
ratios for a large set of stars, including V567~Oph. The radial nature of 
its pulsation was deduced. Hintz \& Joner (1997) and McNamara (1997),
adopting the suggestion by Powell et al. (1990), mentioned the star
in their studies as double-mode variable. Contrary to this, Petersen \& H\o g
(1998) listed V567~Oph as oscillating in fundamental mode only. Musazzi et al.
(1998) went farther, as they refused the secondary periodicity of the star.
It was claimed that the stability of the light curve found by Poretti et al.
(1990) clearly proves the monoperiodic nature. The final argument so far was 
presented by Schwendiman \& Hintz (1999), who gave information on the
existence of a second period of V567~Oph. Unfortunately, in their poster
abstract they did not go into any further details (neither the period value nor
its amplitude was specified).

Other studies dealing at least partly with V567~Oph include Kinman (1998),
who presented an analysis of local space densities of HADS. He noted that 
for four stars (DY~Her, V567~Oph, ZZ~Mic and EH~Lib) it is likely that
they are old disk members instead of belonging to the young galactic disk or
the halo, as other HADS do. Balona \& Evers (1999) discussed the mode
identification of well-observed $\delta$ Scuti stars with the use of 
multicolour photometry. The applied procedure failed to infer
plausible mode identification in three HADS (DY~Her, RS~Gru, V567~Oph). The
single frequency in each case was identified with an $l=1$ g-mode. Balona \&
Evers (1999) concluded that the most likely explanation of their result
is that these stars are evolved radially-pulsating stars outside the range
of the models applied. Although the models show unstable g-modes 
for main-sequence stars, these are unlikely to attain the high 
amplitudes which are observed.

In this paper, we present new Johnson $V$ photometry of V567~Oph which 
revealed the secondary period unambiguously. The observations are described in
Sect.\ 2, while the period analysis is discussed in Sect.\ 3. An interpretation
and possible implications are given in Sect.\ 4.

\section{Observations}

Single-filtered Johnson $V$ observations were carried out at
Szeged Observatory on 15 nights during 2001 and 2002. We used the
0.4m Cassegrain-telescope equipped with an ST--9E CCD camera (512$\times$512
pixels). In order to attain better time resolution we used the
CCD in 2$\times$2 binned mode thus giving an angular resolution of
1.4$^{\prime\prime}$/pixel (the field of view is 6$^\prime\times6^\prime$). 
The exposure time was between 20 and 40 seconds depending on the
weather conditions and the
frames were obtained almost uninterruptedly enabling a very good light curve
coverage during every cycle. The full log of observations is 
given in Table\ 1.

\begin{table}
\begin{center}
\caption{Journal of observations.}
\begin{tabular}{|lrl|}
\hline
Date (yyyy-mm-dd) & No. of points & Length (hours) \\
\hline
2001-07-07 & 329 & 5.3\\
2001-07-08 & 349 & 5.6\\
2001-07-14 & 242 & 4.5\\
2001-07-15 & 205 & 3.4\\
2001-07-20 & 46 & 0.7\\
2001-08-03 & 209 & 3.6\\
2001-08-04 & 266 & 4.7\\
2001-08-12 & 359 & 4.6\\
2001-08-15 & 176 & 2.7\\
2001-08-16 & 112 & 2.2\\
2002-06-18 & 516 & 5.8\\
2002-06-19 & 507 & 5.9\\
2002-06-22 & 563 & 6.1\\
2002-06-26 & 497 & 5.9\\
2002-06-27 & 594 & 6.3\\
\hline
Total: &  4970 & 67.3\\
\hline
\end{tabular}
\end{center}
\end{table}

The data were reduced with standard tasks in IRAF\footnote
{IRAF is distributed by the National Optical Astronomy Observatories,
 which are operated by the
Association of Universities for Research in Astronomy, Inc., under  
cooperative agreement with the National Science Foundation.}. The flat-field
correction utilized sky-flat images taken during the evening or
morning twillight. Differential magnitudes\footnote{Corresponding data
files can be found at CDS via anonymous ftp to {\tt cdsarc.u-strasbg.fr
(130.79.128.5)} or via 
{\tt http://cdsweb.u-strasbg.fr/cgi-bin/qcat?J/A+A/.../...}}
were calculated with aperture photometry using
two comparison stars of similar brightnesses ($C_1$ and $C_4$ in
Powell et al 1990, who gave the following magnitudes: $C_1$ --- 
$y_{\rm mag}=11\fm261$, $b-y=0\fm489$; $C_4$ --- $y_{\rm mag}=11\fm412$,
$b-y=0\fm431$). Since V567~Oph has very similar colour to the applied
comparisons (Table\ 5 in Powell et al. gives a $\langle b-y \rangle=0\fm44$),
the standard transformations are negligible corrections, therefore,
the use of only one filter is not expected to introduce large systematic 
differences.

 The estimated photometric accuracy varied between $\pm0\fm007$
and $\pm0\fm020$ as judged from the rms scatter of the comparison {\it minus}
check data. Although we have tried to reduce the data as careful
as possible, some instrumental drifts of order of a few millimags cannot be
excluded. Due to the small field of view, the variable and comparison 
stars were located at the edge of the frames which may introduces some
additional photometric uncertainty. The nightly mean 
values of the comp--check magnitudes were stable to $\pm0\fm004$ with 
a mean value of $\Delta V=-0\fm193$ with a slight tendency for the 2002
data to be fainter by $\approx0\fm002$ in average. This is below the 
range in which we are interested but might be important when a much
larger and longer dataset becomes available. We also noted 
the 0\fm04 difference between the $\Delta V$ and $\Delta y_{\rm mag}$
values of the comparisons (the latter difference is $-0\fm15$, taking
the magnitudes in Powell et al. 1990), which we attribute to 
the difference between the Str\"omgren $y$ and Johnson $V$ filters.

\section{Period analysis}

\subsection{The $O-C$ diagram}

\begin{table}
\begin{center}
\caption{New times of maximum for V567~Oph ($MJD=HJD-2400000$).
The uncertainties are about $\pm0.0003$ d.}
\begin{tabular}{|llll|}
\hline
$MJD_{\rm max}$ & $MJD_{\rm max}$ & $MJD_{\rm max}$ & $MJD_{\rm max}$\\
\hline
52098.4246 & 52125.3480 & 52444.4262 & 52453.5453\\
52099.4703 & 52126.3818 & 52444.5680 & \\
52105.4578 & 52134.3087 & 52452.3432 & \\
52106.4952 & 52138.3497 & 52452.4945 & \\
\hline
\end{tabular}
\end{center}
\end{table}

First, we wanted to refine the period of V567~Oph by the traditional method of
the $O-C$ diagram. For this, we have collected all times of maximum
from the literature. The data sources were the following works: early
data were taken from the compilation of Powell et al. (1990); since then,
Agerer \& H\"ubscher (1996, 1997, 1998) and Agerer et al. (2001) published
times of maximum quite regularly. We have determined 13 new epochs from
our own light curves with the best phase coverages around maximum
(see Table\ 2.). This was done by fitting low-order (3-5) polynomials
to the parts of the light curves around maximum light. The whole
collection consists of 46 times of maximum between JD 2429785 (1940) and
JD 2452453 (2002). We have calculated the $O-C$ values plotted in the 
top panel of Fig.\ 1 with the following ephemeris (GCVS):

$$Hel.JD_{\rm max}=2438592.4048+0.1495230\cdot E$$

As can be seen in Fig.\ 1, the global trend is linear, and can 
easily be subtracted. The bottom panel shows the residuals, which 
were calculated with the improved period of 
$P=0.149523644\pm0.000000024$ days.
This agrees very well with the period in Powell et al (1990),
as they gave $P=0.149523641\pm0.000000043$ days. We conclude, that the 
global picture of the period constancy is not changed when another
15 years are added to the previous $O-C$ diagram 
by Powell et al. (1990). It means that any relative change of the 
main period (expressed as $(1/P)dP/dt$, Breger \& Pamyatnykh 1998)
is smaller than $2.6\cdot10^{-9}$ year$^{-1}$.

\begin{figure}
\begin{center}
\leavevmode
\psfig{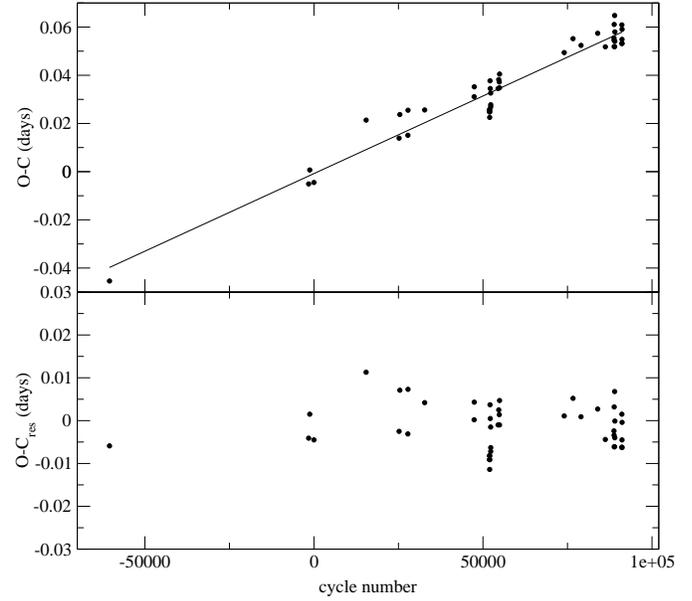}
\caption{The $O-C$ diagram of V567~Oph.}
\end{center}
\label{f1}
\end{figure}

However, as has been already discussed by Powell et al. (1990), 
the $O-C$ diagram shows such large scatter (about $\pm0.01$ days,
i.e. $\pm$15 minutes!) which cannot be observational noise.
Furthermore, our light curves alone showed such cycle-to-cycle
variations which enforced us to reject the assumption of
monoperiodicity. Fortunately, the amount of data enabled 
a more detailed analysis with the Fourier method. 

\subsection{Frequency analysis}

The frequency analysis was performed by means of standard
Fourier-analysis with subsequent prewhitening steps. For this we 
have used Period98 of Sperl (1998) which also includes 
multifrequency least squares fitting of the parameters.
To test our results, we have also re-analysed other datasets in
the literature (Poretti el al. 1990, Powell et al. 1990 and 
Hipparcos Epoch Photometry, ESA 1997).

\begin{figure}
\begin{center}
\leavevmode
\psfig{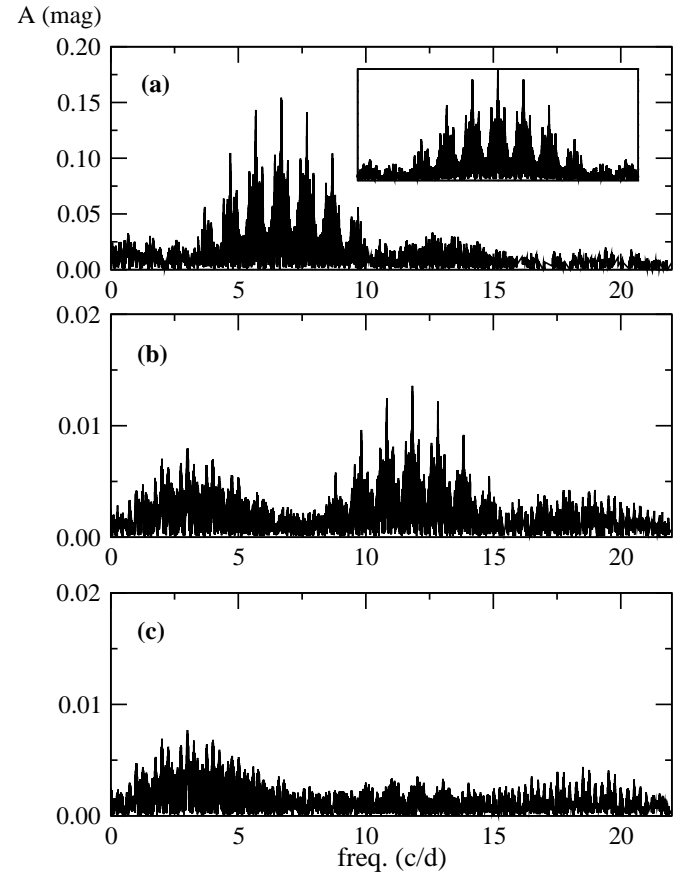}
\caption{Amplitude spectrum of the complete dataset. Insert shows 
the window function (a). After removal of the main period 
and harmonics the secondary period is clearly seen (b). After removal
of the secondary period (c).}
\end{center}
\label{fig2}
\end{figure}

The calculated amplitude spectra are shown in Fig.\ 2. The primary 
peak at $f_1=6.687901$ d$^{-1}$ is in very good agreement with 
the period determined from the $O-C$ analysis (6.6879055 d$^{-1}$).
In every step of the prewhitening procedure we allowed all 
parameters to vary to get the optimal Fourier-fit of the light curve.
The first four prewhitening steps resulted in peaks in the 
subsequent frequency spectra with S/N ratios (Breger et al. 1993)
larger than 10. The fifth step yielded a low-frequency component 
at 3.008 d$^{-1}$ with S/N of 5. However, we did not accept
it because this frequency range is especially sensitive to 
instrumental effects. This has been tested by a Fourier-analysis 
of the magnitude differences between the comparison stars. As expected, 
besides a very low-frequency component (approximately equal to 1/2$\Delta T$,
our time span) we could detect another component
at 2.8 d$^{-1}$ with an amplitude of 3.5 mmag. Therefore, we stopped
the analysis of V567~Oph at the fourth prewhitening step.

As expected from the asymmetric light curve shape, some of the 
frequencies are integer harmonics of the primary one. 
In fact, we have detected $f_1$,
$2f_1$ and $3f_1$ with amplitudes of 0\fm154, 0\fm035
and 0\fm011. The third frequency was the only one independent of $f_1$
at 11.82658 d$^{-1}$ with an amplitude of 0\fm014. Therefore, we
decided to repeat the whole analysis by fixing $f_1$=6.6879055 d$^{-1}$
and its two integer harmonics. After fitting their amplitudes 
and phases only one prewhitening step yielded the secondary frequency
of V567~Oph, hereafter referred to as $f_2$. The parameters of
the adopted frequencies are summarized in Table\ 3, while
the Fourier-fit of the individual light curves are presented 
in Fig.\ 3.

\begin{table}
\begin{center}
\caption{The result of the period analysis.}
\begin{tabular}{|llll|}
\hline
No. & freq. & ampl. & S/N\\
    & (d$^{-1}$) & (mmag) & \\
\hline
$f_1$ & 6.6879055 & 153.1 & 109\\
$2f_1$ & 13.375811 & 34.8 & 31 \\
$3f_1$ & 20.063717 & 10.5 & 11 \\
$f_2$ & 11.82660 & 13.5 & 12.2\\ 
\hline
\end{tabular}
\end{center}
\end{table}

We have separately re-analysed previous data in the literature to
draw some constraints on the stability of the secondary period.
The dataset of Poretti et al. (1990)
allowed the detection of $f_1$ and $2f_1$ and some low-frequency 
noise between 3 and 4 d$^{-1}$ (it has been already noted in that
paper that zero-point shifts affected the data). The Hipparcos Epoch 
Photometry, consisting of 190 data points, yielded only 
$f_1$, though quite accurately (6.687967 d$^{-1}$). The strongest 
pieces of evidence came from the re-analysis of the Powell et al. (1990)
dataset. Besides $f_1$, $2f_1$ and $\sim3f_1$, 
the subsequent prewhitening steps resulted in
$f_2$=11.77 d$^{-1}$ with an amplitude of 0\fm014. Although
$\sim3f_1$ and $f_2$ have low S/N ratio in that dataset, their 
values strongly support our results (note, that the amplitude of the
secondary frequency is very close to ours). Therefore, we conclude that
the light curve of of V567~Oph is doubly periodic and consequently,
the star is a double-mode pulsator.

\begin{figure*}
\begin{center}
\leavevmode
\psfig{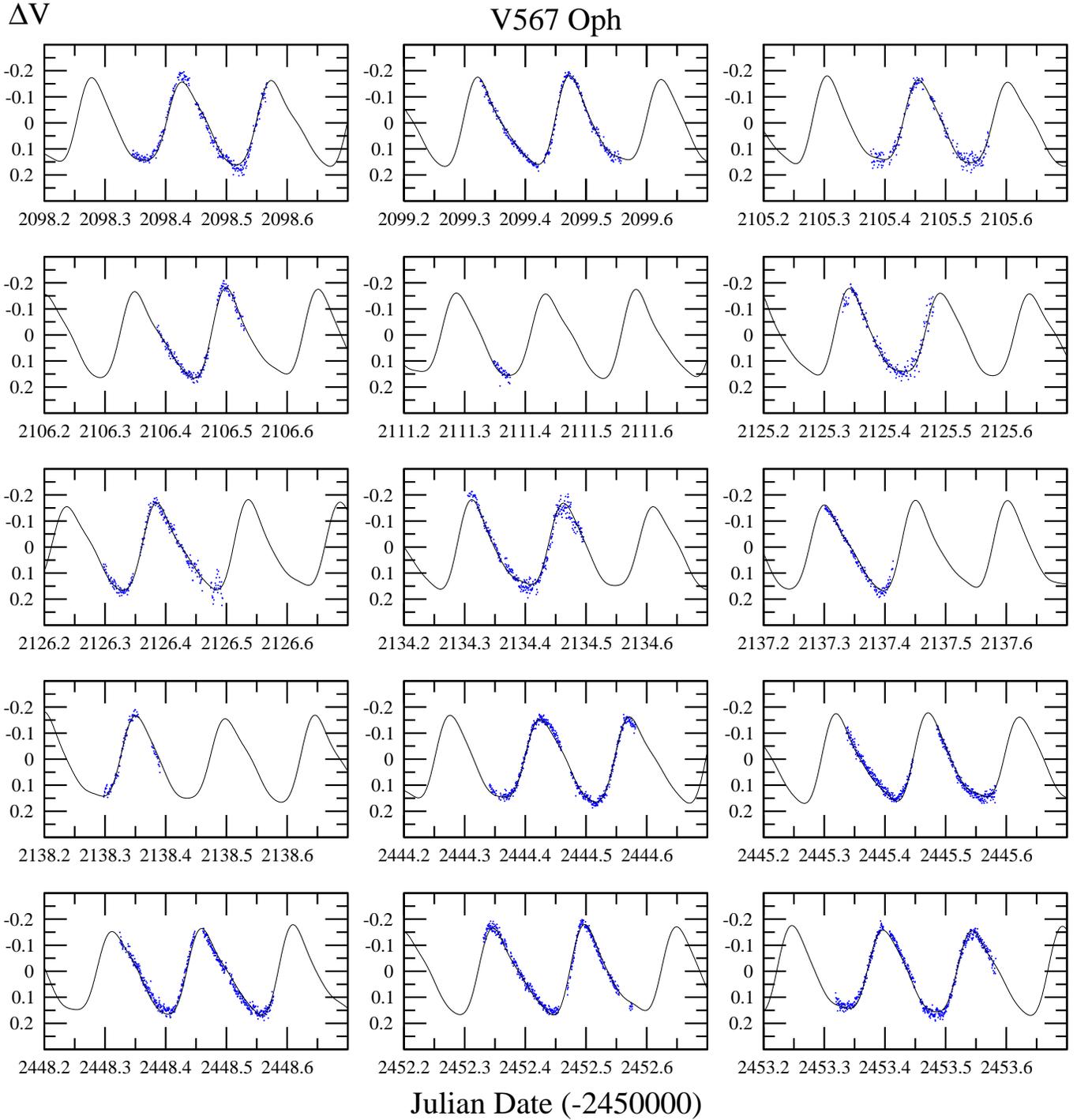}
\caption{The observed individual light curves (small dots) with
the four-component harmonic fit.}
\end{center}
\label{fig3}
\end{figure*}

\section{Discussion}

What can be said about the nature of this double-mode pulsation?
To draw some constraints on mode identification, we have inspected 
the period ratio and $Q$ values. 

The frequency ratio has an intriguing value of $f_1/f_2=0.565$
which is far from the usually found 0.75--0.79 associated with 
radial fundamental and first overtone pulsation (see Petersen \&
Christensen-Dalsgaard (1996) for a parameter study, while Alcock et al. (2000)
for a larger sample of double-mode HADS). As pointed out 
by McNamara (2000), several candidates for higher overtone pulsation 
can be found in $\omega$~Cen, for which recent linear nonadiabatic
models (Gilliland et al. 1998) predict $f_0/f_2=0.63$ and $f_0/f_3=0.53$.
However, neither of these two values fits the observed one which  
is fairly accurate (its error is less than $\pm$0.001). From the
observational
point of view, it is interesting that there is no other double-mode 
HADS with similar frequency ratio. Therefore, we searched the literature 
to find similar stars among the lower amplitude $\delta$ Scuti stars.
Close resemblance was found for AN~Lyn (0.565, Rodr\'\i guez et al. 1997, 
Zhou 2002), V663~Cas (0.591, Mantegazza \& Poretti 1990) and 63~Her
(0.564, Breger et al. 1994). In all cases the authors arrived to
the conclusion that a mixture of radial and non-radial modes
is needed to explain the ``non-standard'' frequency ratios. 
Adopting their argumentations, it is reasonable to accept this 
consideration in the case of V567~Oph, too. Recent theoretical models
also support this idea (Bono et al. 1997, Gilliland et al. 1998).

We have calculated the pulsation constant $Q$ for both frequencies with the 
formula

$$\log Q=-6.456+\log P+0.5 \log g + 0.1 M_{\rm bol} + \log T_{\rm eff}$$

\noindent in terms of four observables (Breger et al. 1993). The physical 
parameters of V567~Oph were determined 
by Powell et al. (1990), McNamara (1997) and Balona \& Evers (1999).
We list the corresponding $Q$ values in Table\ 4.

\begin{table}
\begin{center}
\caption{$Q$ values from various physical parameter 
determinations}
\begin{tabular}{|llllll|}
\hline
Ref. & $\log g$ & $M_{\rm bol}$ & $\log T_{\rm eff}$ & $Q_1$ & $Q_2$ \\
\hline
(1) & 3.74 & 1\fm1 & 3.87 & 0.037 & 0.021\\
(2) & 3.76 & 0\fm85 & 3.87 & 0.035 & 0.020\\
(3) & 3.88 & 1\fm40 & 3.87 & 0.047 & 0.027\\
(3) & 3.26 & 1\fm34 & 3.90 & 0.024 & 0.013\\
\hline
\end{tabular}
\end{center}
References: (1) Powell et al. (1990), (2) McNamara (1997), (3) Balona 
\& Evers (1999). 
\end{table}

While the first two sets of parameters are in good 
agreement, the last two are fairly contradictory. 
Balona \& Evers (1999) used different calibrations. For V567~Oph, they 
applied calibrations by Balona (1994) and Moon \& Dworetsky (1985) --
the latter one produced that deviant $\log g$=3.26 resulting in 
hardly acceptable $Q$ values. If we keep the first set of parameters
of Balona \& Evers,
it still results in uncomfortably high
$Q_1$. However, this is likely caused by a systematic error in
their Str\"omgren calibrations, an error that
was pointed out by Rodr\'\i guez \& Breger (2001).
From a comparison of photometric and geometric parallaxes they found
that for slowly rotating stars Str\"omgren calibrations underestimate 
the absolute magnitude by 0\fm5 in average. Since V567~Oph is 
a typical slowly rotating HADS ($v\sin i<18$ km~s$^{-1}$, McNamara 1985),
its absolute magnitude is likely to be affected by this systematic error.
A correction by $\Delta M_{\rm V}=0\fm5$ shifts $Q_1$ to 0.036 
and $Q_2$ to 0.020,
being in good agreement with the other two calibrations. 
 
Assuming 20\% uncertainty in $Q$ values (that is $\pm0.007$ and $\pm0.004$
for $Q_1$ and $Q_2$, respectively) we conclude that the dominant 
period indeed corresponds to the radial fundamental mode, as 
assumed for the overwhelming majority of HADS. On the 
other hand, the secondary period can be identified with a non-radial
mode of radial order $n=2$ or 3 (Bono et al. 1997, Gilliland et al. 1998).
Presently nothing can be specified unambiguously about the non-radial
degree $l$, for which the simpliest assumption is 1 or 2.

In order to perform more secure mode identification, accurate multicolour 
photometry would be of great importance. Unfortunately, the 
relative faintness of the star does not make it favourable
target object for photoelectric photometry. Therefore, CCD observers 
with proper instrumentation are expected to gain more insights into
the peculiar pulsation pattern of V567~Oph by taking follow-up 
observations of this interesting HADS.

\begin{acknowledgements}
This work has been supported by the Hungarian OTKA
Grants \#T032258 and \#T034615, the ``Bolyai J\'anos'' Research 
Scholarship to LLK from the Hungarian Academy of Sciences, the
Hungarian E\"otv\"os Fellowship to LLK, FKFP Grant 0010/2001,
Pro Renovanda Cultura Hungariae Foundation
and Szeged Observatory Foundation. LLK thanks the kind hospitality 
of the Instituto de Astrof\'\i sica de Andaluc\'\i a, where 
the analysis has been finished. Useful comments and suggestions
by E. Rodr\'\i guez are also acknowledged.
The NASA ADS Abstract Service was used to access data and
references. This research has made use of the SIMBAD database, operated at
CDS-Strasbourg, France.
\end{acknowledgements}

\end{document}